\documentclass[aps,prl,twocolumn,groupedaddress]{revtex4}
\usepackage{graphicx}
\usepackage{braket}
\usepackage{amsmath}
\usepackage{mathrsfs}
\usepackage{color}
\usepackage{ulem}

\begin{document}


\title{Two-pulse field-free orientation reveals anisotropy of molecular shape resonance}


\author{P. M. Kraus}
\affiliation{Laboratorium f\"ur Physikalische Chemie, ETH Z\"urich, 8093 Z\"urich, Switzerland}
\author{D. Baykusheva}
\affiliation{Laboratorium f\"ur Physikalische Chemie, ETH Z\"urich, 8093 Z\"urich, Switzerland}
\author{H. J. W\"orner}
\email[]{woerner@phys.chem.ethz.ch}
\homepage[]{www.atto.ethz.ch}
\affiliation{Laboratorium f\"ur Physikalische Chemie, ETH Z\"urich, 8093 Z\"urich, Switzerland}


\date{\today}

\begin{abstract}

We report the observation of macroscopic field-free orientation, i.e. more than 73 \% of CO molecules pointing in the same direction. This is achieved through an all-optical scheme operating at high particle densities ($>10^{17}$cm$^{-3}$) that combines a one-color ($\omega$) and a two-color ($\omega + 2\omega$) non-resonant femtosecond laser pulses. We show that the achieved orientation solely relies on the hyperpolarizability interaction as opposed to an ionization-depletion mechanism thus opening a wide range of applications. The achieved strong orientation enables us to reveal the molecular-frame anisotropies of the photorecombination amplitudes and phases caused by a shape resonance. The resonance appears as a local maximum in the even-harmonic emission around 28~eV. In contrast, the odd-harmonic emission is suppressed in this spectral region through the combined effects of an asymmetric photorecombination phase and a sub-cycle Stark effect, generic for polar molecules, that we experimentally identify.  
\end{abstract}



\maketitle

Techniques for fixing molecules in space are invaluable tools for a broad range of experiments in ultrafast science \cite{stapelfeldt03a,rakitzis04a}. The availability of transiently aligned molecular samples has particularly advanced strong-field and attosecond spectroscopies, providing new insights into the electronic structure of molecules and its temporal evolution \cite{itatani04a,meckel08a,woerner10b,woerner11c,vozzi11a}. High-harmonic spectroscopy (HHS) provides a new access to the rich structures of photoionization continua, such as Cooper minima \cite{woerner09a,bertrand12a,wong13a,ren13a} and shape resonances \cite{shiner11a,rupenyan12a,ren13a}. The investigation of the inherent structural and dynamical anisotropies of polar molecules has however been prevented by the difficulty of orienting molecules. Interesting phenomena tied to polar molecules include the predicted recombination-site dependence of structural minima \cite{rupenyan13a,rupenyan13b} and attosecond charge migration \cite{cederbaum99a,breidbach03a,kraus13c} triggered by strong-field ionization. Here, we demonstrate a protocol for molecular orientation which achieves macroscopic field-free orientation and exploit this progress to probe the anisotropy of photorecombination dipole moments at a molecular shape resonance.

Successful approaches to laser-induced molecular orientation include the combination of an electrostatic field with a rapidly turned-off laser field \cite{goban08a}, alignment in combination with quantum-state selection and a weak dc-field \cite{holmegaard09a,holmegaard10a,vrakking97a,ghafur09a}, adiabatic \cite{oda10a} and impulsive two-color orientation \cite{de09a,kraus12c,frumker12a}. All of these techniques are subject to substantial limitations: The presence of electric fields may alter the electronic structure of the molecule, its photo-induced dynamics or the subsequent probing process. The low particle densities available after quantum-state selection make the application of such techniques to high-harmonic and attosecond spectroscopies challenging to impossible. The two-color scheme \cite{vrakking97a,de09a}, recently applied to HHS \cite{kraus12c,frumker12a}, relies on an ionization-depletion mechanism \cite{spanner12a} and thus ties the achievable degree of orientation to the ionization fraction of the sample. This fact does not only limit orientation to modest degrees but it also makes charged-particle detection difficult.

In this letter, we describe a technique of molecular orientation that overcomes all of the above-mentioned limitations. By applying a one-color alignment pulse (170~fs duration, (5$\pm$1)$\times$10$^{13}$ W/cm$^2$) followed by a two-color orientation pulse (150 fs, (6$\pm$1)$\times$10$^{13}$ W/cm$^2$ ) to CO molecules in a supersonic expansion (Fig.~1(a)), as theoretically proposed in \cite{zhang11a,tehini12a}, we achieve macroscopic orientation, i.e. more than 73 \% of the molecules pointing in the same direction, under completely field-free conditions. 
Details of the experimental setup have been given elsewhere \cite{kraus12c,kraus13a}. 
The degrees of orientation obtained in this study rival those obtained after quantum-state selection \cite{ghafur09a} with a $\sim $10$^{6}$-fold increase in particle density. Moreover, we show through calculations that the present technique solely relies on the interaction of the laser field with the molecular hyperpolarizability thus circumventing the need to ionize major parts of the sample \cite{spanner12a}.

We exploit this technique to showcase the complementary encoding of electronic-structure information in the even and odd high-harmonic spectra of oriented molecules. We identify a shape resonance in CO around 28~eV that manifests itself as a pronounced local maximum in the intensity of the even orders while the odd orders are suppressed in the same region. Quantitative modelling shows that this remarkable observation is the consequence of a $\sim\pi$-phase shift of high-harmonic radiation emitted from the two ends of the molecule from the 15$^{\rm th}$ to the 22$^{\rm nd}$ order (H15-H22). This phase shift is caused by the combined effects of the recombination phase and a sub-cycle Stark effect, generic for polar molecules. The present study thus opens the field of side-dependent attosecond photorecombination delays \cite{schultze10a,kluender11a} to polar molecules.

\begin{figure}
\includegraphics[width=0.5\textwidth]{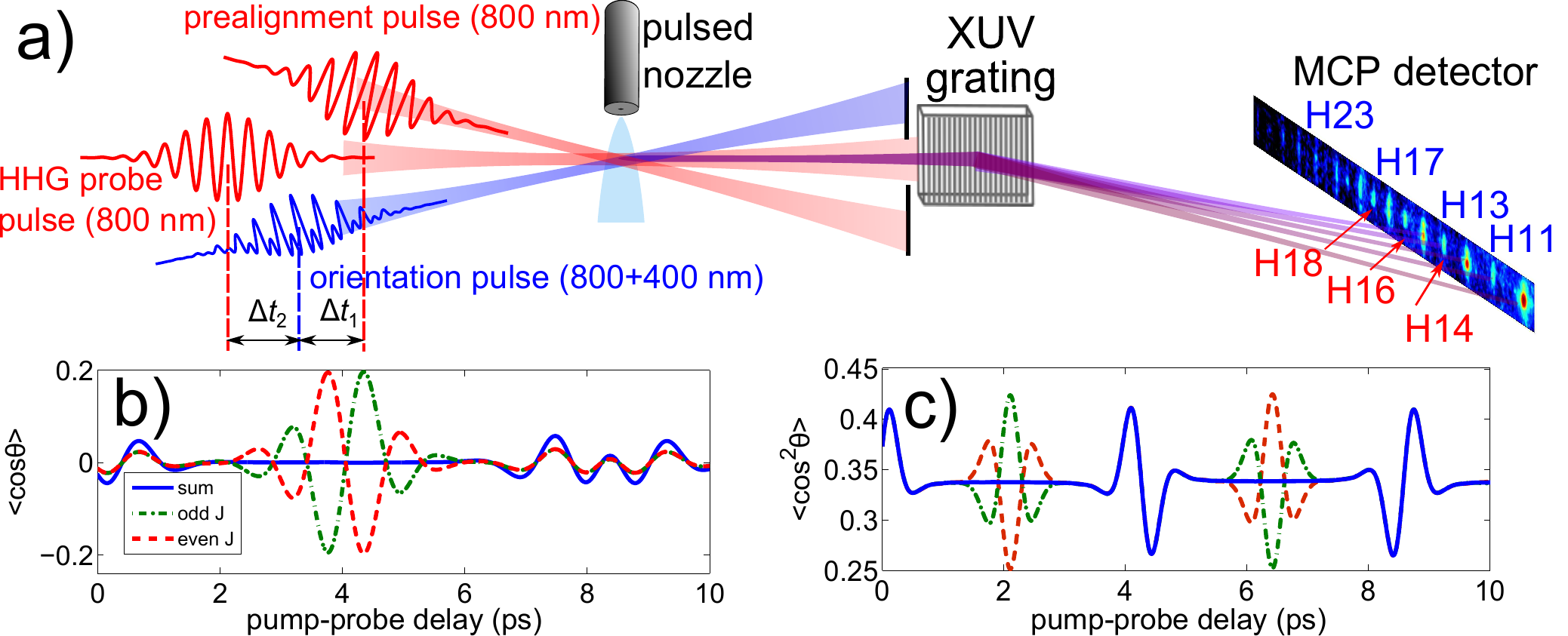}
\caption{(a) Experimental setup for high-harmonic spectroscopy of molecules oriented with the two-pulse scheme. Calculated time evolution of (b) $\braket{\cos\theta}$ for CO oriented by a single two-color laser pulse and of (c) $\braket{\cos^2\theta}$ for CO aligned by a single one-color laser pulse. The contributions of rotational levels with even or odd angular momentum quantum number (\textit{J}) and their sum are shown separately in both panels. \label{}}
\end{figure}

Figures~1(b) and 1(c) illustrate the concept of our experiment through calculated values of orientation and alignment parameters for CO molecules ($B=1.93128~\textnormal{cm}^{-1}$, $T_{\textnormal{rot}}=8.63~\textnormal{ps}$, $I_p= 14.014~\textnormal{eV}$ \cite{nistCO}). The two-color orientation pulse alone creates only weak net orientation at the full rotational revival ($T_{\textnormal{rot}}$, see Fig.~1(b)), whereas it generates strong orientation in the isolated contributions of the even- and odd-\textit{J} (rotational quantum number) levels at the rotational half-revival (around $T_{\textnormal{rot}}$/2). These contributions however interfere destructively yielding no net orientation. The application of a one-color pulse $\Delta t_1=T_{\textnormal{rot}}$/4 before the two-color pulse, leads to the alignment (anti-alignment) of the odd- (even-) $J$ contributions (Fig.~1(c)). The two-color pulse can then selectively orient the pre-aligned odd-$J$ contributions and create strong net orientation at the rotational half-revival. Fine-tuning of the exact delays even enables constructive interference to be achieved between odd- and even-$J$ contributions to orientation \cite{kraus14a}.

\begin{figure}
\includegraphics[width=0.5\textwidth]{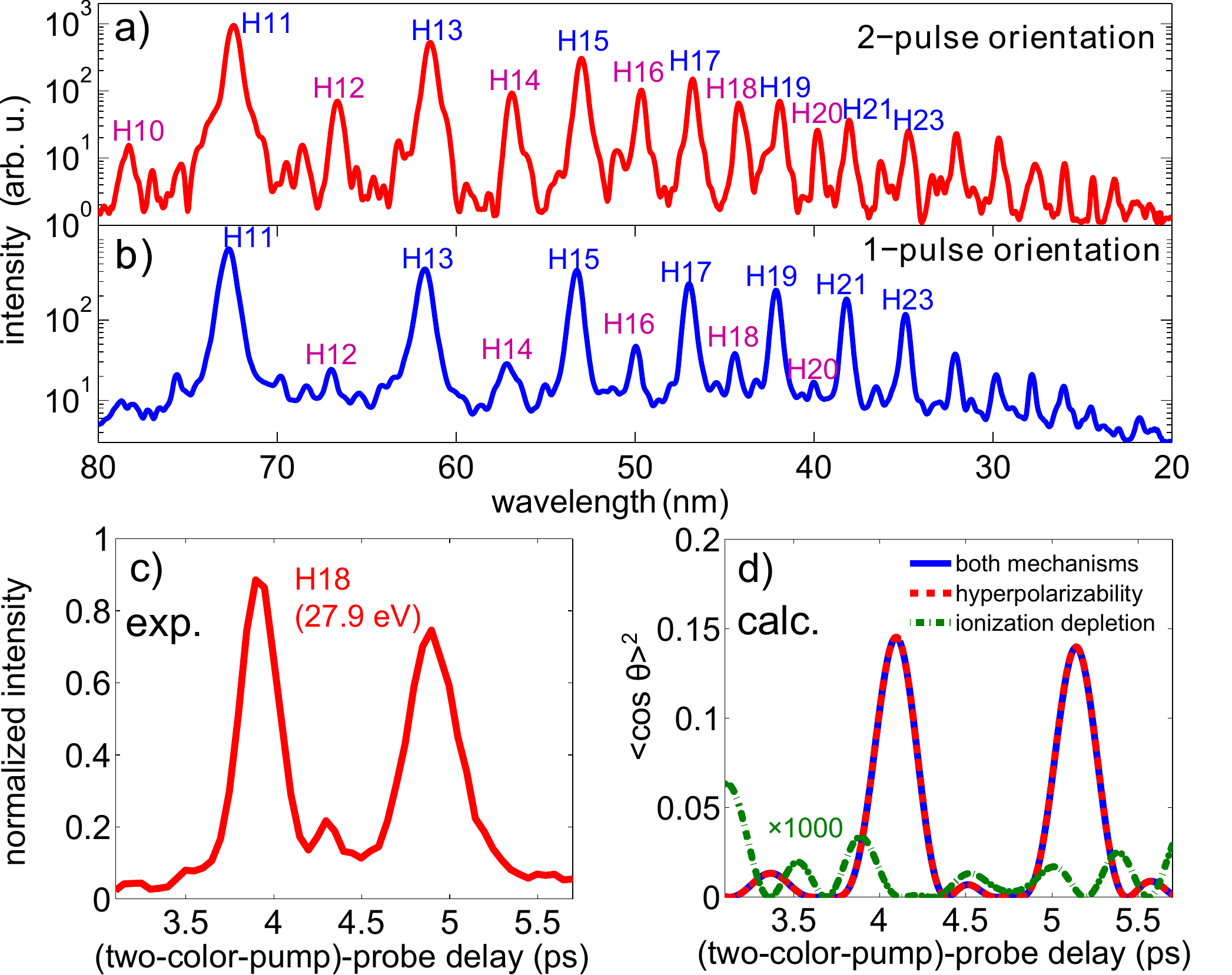}
\caption{(a) High-harmonic spectrum of CO ((2.2$\pm$0.1)$\times$10$^{14}$ W/cm$^2$) oriented by the two-pulse scheme. (b) High-harmonic spectrum of CO oriented by the one-pulse scheme recorded under conditions otherwise identical to the upper spectrum. (c) Variation of the intensity of a representative even harmonic (H18) with pump-probe delay for the two-pulse scheme. The intensity was normalized to the intensity of the next higher odd harmonic (H19) at the delay of maximal orientation. (d) Calculated evolution of $\braket{\cos\theta}^2$ including only the hyperpolarizability mechanism of molecular orientation, only the ionization depletion mechanism or both mechanisms.\label{}}
\end{figure}

Figure~2(a) shows the high-harmonic spectrum of CO molecules oriented with the two-pulse scheme. The delay between the two-color orientation and alignment pulses was set to $\Delta t_1=2.25$~ps and the delay between the orientation and HHG pulses was $\Delta t_2=3.90$~ps. The emission of intense even-harmonic orders, with H18 and H19 displaying comparable intensities, indicates macroscopic orientation of the sample. The even-harmonic emission is dramatically enhanced as compared to orientation by a single two-color laser pulse shown in Fig.~2(b). This latter spectrum was recorded at a pump-probe delay of 8.85~ps and shows even-harmonic emission that is weaker by a factor of $\sim$ 9, as compared to the spectrum obtained with the two-pulse scheme (Fig.~2(a)). In both experiments, the time delays were optimized to obtain maximal orientation whereas the other experimental settings were kept identical. The evolution of the even-harmonic emission as a function of $\Delta t_2$ is shown in Fig.~2(c) for a representative harmonic order (H18, 27.9~eV). 

We calculate the wave-packet dynamics by solving the time-dependent Schr\"{o}dinger equation (TDSE) as previously described in e.g. \cite{zhang11a,tehini12a} using the hyperpolarizability interaction and additionally incorporate the ionization-depletion mechanism \cite{spanner12a} with details given in Ref. \cite{kraus14a}. Figure 2(d) shows the calculated value of $\braket{\cos\theta}^2$ which nicely reproduces the temporal evolution of the even-harmonic signal. Strikingly, orientation is found to exclusively result from the hyperpolarizability interaction (dashed red line) whereas the contribution of the ionization-depletion mechanism (dash-dotted green line) is negligible. The present two-pulse scheme for orientation is thus the first field-free scheme which orients molecules without ionizing a substantial fraction of the sample. Hence, it is a valuable tool for all probe techniques that suffer from ionization of the sample by pump pulses, such as photoelectron spectroscopies.  

We now study the encoding of electronic-structure information in the even- and odd-harmonic spectra of polar molecules and quantify the degree of orientation. A local maximum is observed in the even-harmonic spectrum at H18 (27.9~eV) (Fig.~2(a)), which is even more pronounced in the ratio of the even harmonics divided by the mean value of their two adjacent odd harmonic orders (the even-to-odd ratio, red dots in Fig.~3(a)). This local maximum is independent of the probe-pulse intensity. Contrary to the even-harmonic spectrum, the intensities of the odd harmonic orders are continuously decreasing over the spectral range where the even harmonic emission maximizes. The analysis of the temporal modulation of the odd harmonic intensities further shows that the latter are strongly suppressed in the same spectral region (H17-cutoff, 26.3-48 eV) when CO is aligned parallel to the HHG probe pulse. This observation is illustrated in Figs.~3(d) and (e) which show results from a one-color-pump HHG-probe experiment under conditions otherwise identical to those described above. The calculated degree of axis alignment measured by $\braket{\cos^2\theta}$ (Fig.~3(c)) maximizes around 8.8~ps. At this delay, the intensity of the low harmonic orders (H9-H15, Fig. 3(d)) maximizes as well, whereas the intensity of the higher harmonics (H17-cutoff, Fig. 3(e)) minimizes \cite{frumker12b}.

\begin{figure}
\includegraphics[width=0.5\textwidth]{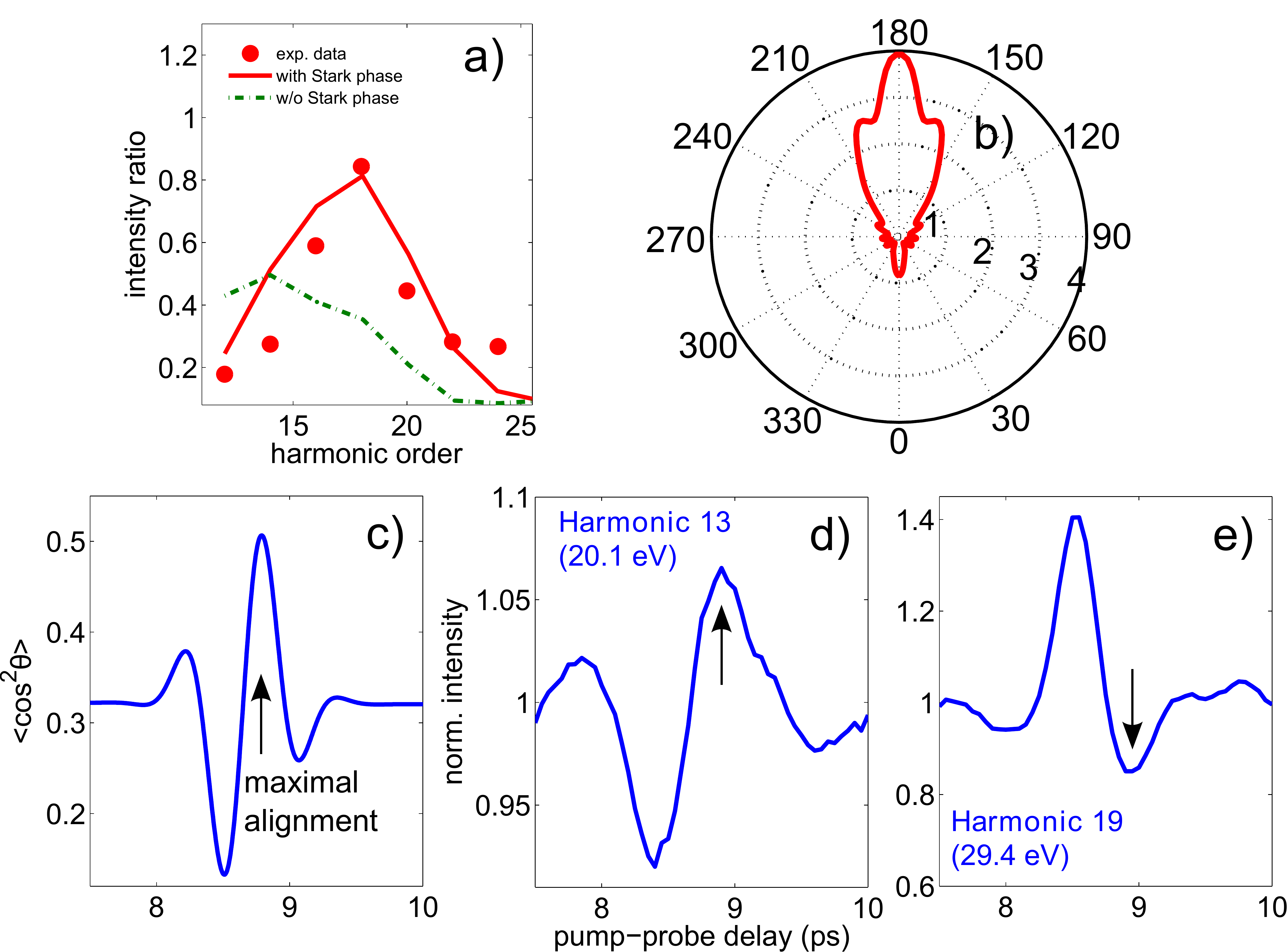}
\caption{(a) Experimental and simulated even-to-odd ratio. The ratio was simulated with and without taking the Stark phase into account. (b) Oriented axis distribution (c) Calculated evolution of $\braket{\cos^2\theta}$. (d) and (e) Measured evolution of the odd harmonic intensities as function of the pump-probe. The time of maximal alignment is marked with an arrow in the panels (c)-(e). \label{}}
\end{figure}

To explain these observations, we simulate the high-harmonic spectra of oriented CO molecules. We calculate the single-molecule induced dipole moment as product of the recombining electron wave packet -- obtained as the product of the ionization amplitude $\sqrt{I_\mathrm{ion,i}(\theta)}$ and a propagation factor $a_\mathrm{prop,i}(\Omega,\theta)$ -- and the photorecombination dipole moment $d_\mathrm{rec,i}(\Omega,\theta)$. $\theta$ represents the polar angle between the molecular axis and the laser field and $\Omega$ stands for the angular frequency of the emitted radiation. The induced dipole moment is coherently averaged over the molecular axis distribution $A(\theta)$ obtained from a TDSE calculation and summed over the channels $i$ corresponding to ionization from the highest-occupied molecular orbital (HOMO) and HOMO-1 \cite{rupenyan12a,rupenyan13a}:
\begin{equation}
\begin{split}
\ d(\Omega) \propto &\sum_i \int_{0}^{\pi} A(\theta) \sqrt{I_\mathrm{ion,i}(\theta)}a_\mathrm{prop,i}(\Omega,\theta)
\\ & \times d_\mathrm{rec,i}(\Omega,\theta) \sin\theta \, \mathrm{d}\theta.
\label{eq:HHGdip}
\end{split}
\end{equation}

We refer here to orbitals rather than electronic states of the cation because the X $^2\Sigma^+$ and A $^2\Pi$ states of CO$^+$ are well described by single-hole configurations of the neutral molecule. We obtain the ionization rate for the HOMO of CO from recent theoretical work \cite{spanner12a} and calculate the ionization rate of HOMO-1 using the partial Fourier-transform method \cite{murray11a}. Their relative rates are chosen to reproduce the results of Ref. \cite{wu12a} and scaled to the intensity applied in our study. We obtain the complex amplitude of $a_\mathrm{prop,i}(\Omega,\theta)$ numerically using the strong-field approximation \cite{yakovlev07a}. Only the phase of $a_\mathrm{prop,i}(\Omega,\theta)$ is angle dependent as a consequence of a linear sub-cycle Stark effect of the ionized orbital \cite{etches10a} which can be expressed as scalar product of the permanent dipole moment of the orbital and the return velocity of the electron \cite{etches12a}. The Stark phase is thus independent of the intensity and wavelength of the driving laser field and only depends on the emitted photon energy. The photorecombination dipole moments are obtained through \textit{ab initio} quantum scattering calculations using \textsc{epolyscat} \cite{gianturco94a,natalense99a} with a polarized valence-triple-zeta (pVTZ) basis set.

From our experimental data we obtain an even-to-odd ratio of 0.44 determined as an average over the harmonic orders 12 to 24 (H12-H24). This allows us to roughly estimate $\eta=\frac{n_\textnormal{up}-n_\textnormal{down}}{n_\textnormal{up}+n_\textnormal{down}}=0.66$, i.e. $\zeta=\frac{n_\textnormal{up}}{n_\textnormal{tot}}=0.83$, because the spectrally-averaged even-to-odd ratio is proportional to $\eta^2$. This is consistent with both the experimentally observed enhancement of the even orders by a factor of 9 and the corresponding improvement in the degree of orientation $\eta$ by a factor of 3 compared to Ref. \cite{frumker12a}. Since the odd (even) harmonic orders correspond to the sum (difference) of the electric fields emitted from the two sides of the molecule \cite{kraus12c,frumker12b}, we simulate the odd (even) harmonic spectra using a symmetrized, $A_{\mathrm{sy}}(\theta)=(A(\theta)+A(\pi-\theta))/2$ (anti-symmetrized, $A_{\mathrm{as}}(\theta)=A(\theta)-A(\pi-\theta)$/2) version of the normalized axis distribution $A(\theta)$ obtained from the TDSE calculation shown in Fig.~3(b) ($\zeta=0.73$, $\braket{\cos\theta}=0.38$). The calculated and measured even-to-odd ratios are shown together in Fig.~3(a). The full theory (solid red line) is in very good agreement with the measured even-to-odd ratio (red dots). From these two independent approaches to estimating the degree of orientation, we conclude $0.73 \leq \zeta \leq 0.83$. 

\begin{figure}
\includegraphics[width=0.5\textwidth]{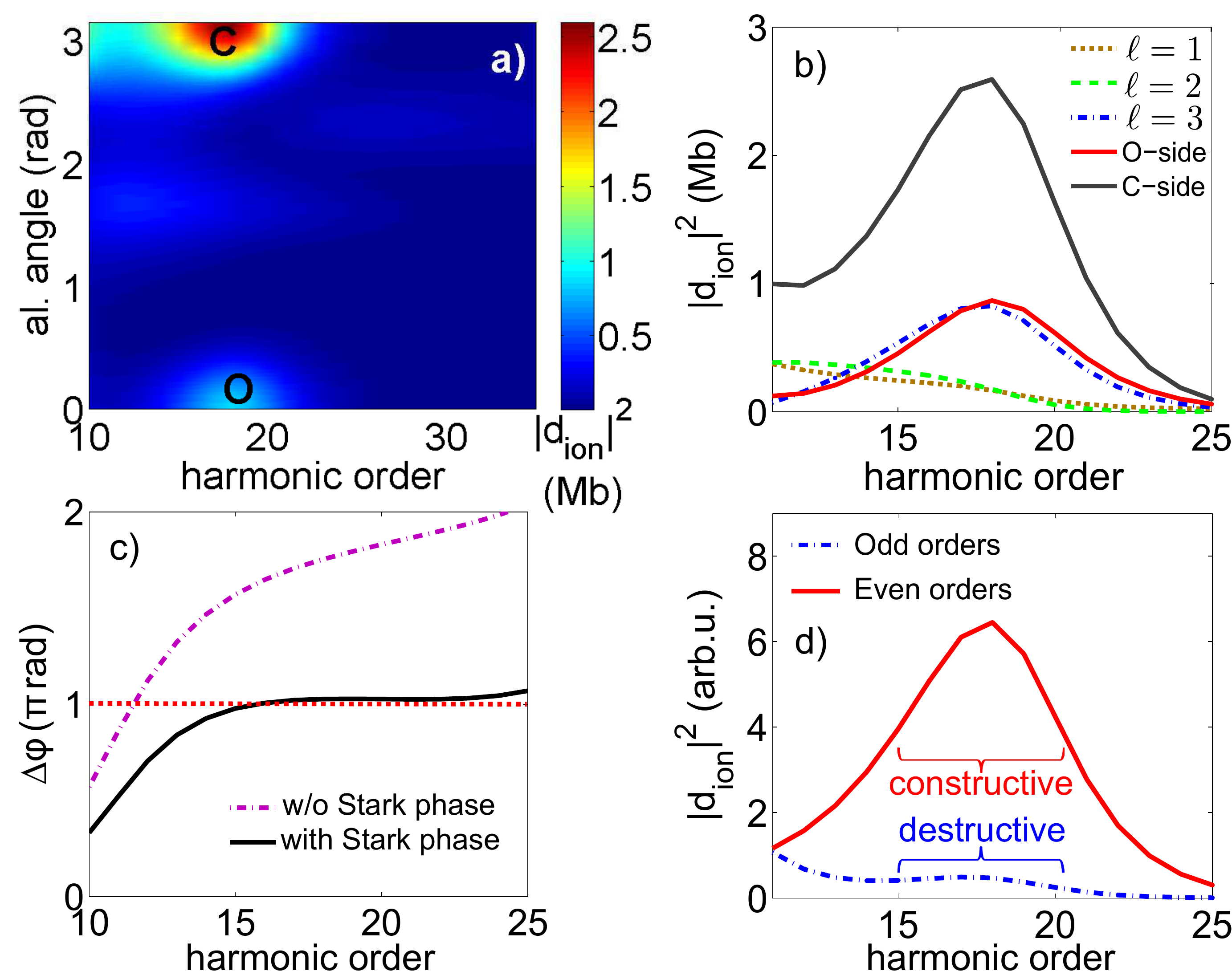}
\caption{(a) Molecular-frame photoionization cross section for the HOMO of CO as function of the alignment angle and the harmonic order. (b)~Decomposition of the photoionization cross section into partial waves ($\ell=1,2,3$) for parallel alignment of the molecule and differential photoionization cross sections for electron emission from the O-side (0$^\circ$) or the C-side (180$^\circ$) of the molecule. (c) Phase difference $\Delta\varphi$ of high-harmonic emission from the C- or O-sides of the molecule. (d) Squared magnitude of the sum and difference of the recombination dipoles including the Stark-phase contributions, reflecting the envelope of odd- and even-harmonic spectra, respectively.\label{}}
\end{figure}

We now rationalize the maximum in the even-harmonic emission and the concomitant suppression of odd-harmonic emission. An excellent agreement between theory and experiment is achieved when both HOMO and HOMO-1 are taken into account (red line in Fig.~3(a)). Neglecting the contributions of the Stark effect shifts the maximum from H18 to H14 and substantially deteriorates the overall agreement (green dashed line), demonstrating the importance of the Stark effect. However, the local maximum at 28~eV is unchanged when HOMO-1 is excluded from the calculation, which allows us to concentrate on HOMO. Figure 4(a) displays the calculated differential photoionization cross section of the HOMO of CO for photoemission along the positive $z$-axis defined by the polarization of the ionizing radiation. This cross section displays a pronounced maximum around H18 (27.9~eV). The maximum occurs both when the oxygen (0$^\circ$ alignment angle) or the carbon (180$^\circ$) atom point in the positive $z$ direction, with the latter being more pronounced. We demonstrate the origin of this maximum by decomposing the cross section into the contributions from individual partial waves (Fig.~4(b)). The prominent maximum in the $\ell=3$ contribution (dash-dotted blue line) arises from a centrifugal barrier in the effective potential determined by the angular momentum $\ell$ of the continuum electron, which causes the latter to be trapped in a quasibound state. This phenomenon is commonly referred to as ''shape resonance'' and has been observed at 23.6 eV in the partial photoionization cross section of CO pertaining to the X $^2\Sigma^+$ state of CO$^+$ \cite{plummer77a}. The shape resonance observed in our experiment lies 4.3 eV higher compared to the shape resonance observed in photoionization. This shift is mainly caused by the Stark effect as shown in Fig. 3a. The observed width of the shape resonance (7.5-8 eV) is very similar to that observed in photoionization (8 eV).

We now demonstrate an application of HHS to probing the hitherto inaccessible anisotropy of molecular shape resonances. The shape resonance causes unequal enhancements of the differential photoionization cross section on the two sides of the molecule (Fig. 4~(b)). On the oxygen end, the partial-wave contributiosn with $\ell=1$ and $\ell=2$ interfere destructively, such that the energy-dependence of the $\ell=3$ contribution is recovered. The carbon side shows an enhanced cross section caused by a constructive interference of the $\ell=1-3$ contributions. This demonstrates that HHS of oriented polar molecules directly probes the energy- and angle-dependent contributions of isolated partial waves to the photoionization cross section. The shape resonance further causes a pronounced energy dependence in the phase of the photoionization matrix elements. This is illustrated in Fig. 4~(c) which shows the difference of the photorecombination dipole phases between C- and O-sides (dash-dotted line). The total phase difference between high-harmonic emission from the C- and O-sides (solid black line) is obtained by adding the Stark-phase contribution \cite{etches10a} to the difference of the photorecombination phases and is found to be nearly constant around $\Delta\varphi=\pi\pm0.1~\textnormal{rad}$ in the range of H15-H22 (22-34 eV). Figure~4(d) illustrates the effect of this phase difference on the high-harmonic spectrum. The squared magnitudes of the coherent sum (dash-dotted blue line) and difference (solid red line) of the recombination dipoles from the C- and O-sides including the Stark phase are shown, reflecting the intensity envelopes of the odd and even harmonic orders. The $\sim\pi$ phase difference in the range H15-H22 causes a strong suppression of the odd harmonics in this region, whereas the even harmonics display a maximum at H18. 

These results explain why the even-harmonic emission maximizes around the photon energy of the shape resonance (Fig. 3(a)), whereas the odd-harmonic emission from aligned molecules is suppressed in the same spectral range (Fig. 3(e)). High-harmonic spectroscopy is thus shown to be sensitive to the anisotropic dipole phase caused by a molecular shape resonance. Our analysis further reveals the striking importance of the sub-cycle Stark effect. If the latter is neglected, the dipole phase difference reaches $\pi$ around H12 (Fig. 4(c)), explaining why the maximum in the even-harmonic spectrum is shifted to lower photon energies compared to the experiment when the Stark effect is neglected (dash-dotted green line in Fig. 3(a)).

In conclusion, we have demonstrated an all-optical technique that creates macroscopic degrees of field-free molecular orientation, i.e. more than 73 \% of the molecules pointing in one direction. This technique creates orientation solely through the hyperpolarizability interaction and overcomes the limitations of previous techniques such as low degrees of orientation, low particle densities or ionization of the sample. This progress enables us to characterize the anisotropy of complex photorecombination dipole moments at a molecular shape resonance. These anisotropies were inaccessible to previous synchrotron studies which only provided angle-averaged photoionization cross sections. In particular, the phases of photoionization dipole moments become accessible to HHS and contain valuable information on the electronic structure and dynamics. The pronounced asymmetry of the recombination dipole phase combined with a sub-cycle Stark effect, generic for polar molecules, causes the high-harmonic emission from the two sides of the molecule to be $\sim\pi$ radians out of phase over an extended spectral range. This leads to the appearance of a spectral maximum only in the even-harmonic spectrum while the odd harmonics are suppressed in the same spectral region.
The antagonistic encoding of spectral features, such as photorecombination resonances or Cooper minima, is expected to be generic for even- and odd-harmonic spectra as a consequence of their complementary dependence on the emissions from opposite recollision sites. The present results make macroscopic orientation available to high-harmonic and attosecond spectroscopies and open new directions of research such as the study of electronic dynamics in bound states, continuum resonances or photorecombination delays in polar molecules.

\end{document}